\documentclass{sig-alternate}
\usepackage{mathptmx} 
\newcommand{\ignore}[1]{100}

\usepackage{fancyhdr}
\usepackage[normalem]{ulem}
\usepackage[hyphens]{url}
\usepackage{color}
\usepackage{microtype}
\usepackage{authblk}

\usepackage[bookmarks=true,breaklinks=true,letterpaper=true,colorlinks,linkcolor=black,citecolor=blue,urlcolor=black]{hyperref}

\usepackage{enumitem}
\usepackage{lipsum}
\setlist[enumerate]{wide=\parindent}

\usepackage{graphicx}
\usepackage{color,soul}
\definecolor{lightgreen}{RGB}{195, 233, 211}
\sethlcolor{lightgreen}

\soulregister\cite7
\soulregister\ref7
\soulregister\pageref7

\makeatletter
\newcommand{\printfnsymbol}[1]{%
  \textsuperscript{\@fnsymbol{#1}}%
}
\makeatother

\title{MARS: Memory Aware Reordered Source}

\author[1]{Ishwar Bhati \thanks{The author contributed to the work while working at Intel Labs}  \protect\phantom{\footnotesize} }
\author[2]{Udit Dhawan}
\author[1]{Jayesh Gaur }
\author[1]{Sreenivas Subramoney}
\author[1]{Hong Wang }

\affil[1]{Intel Labs}
\affil[2]{Turtle Shell Technologies}
\affil[ ]{\textit {\{ishwar.s.bhati@intel.com\}}}

\begin{document}
\maketitle
\pagestyle{plain}
\begin{abstract}
\fontsize{9.76}{10.76}\selectfont
Memory bandwidth is critical in today's high performance computing systems. The bandwidth is particularly paramount for GPU workloads such as 3D Gaming, Imaging and Perceptual Computing, GPGPU due to their data-intensive nature. As the number of threads and data streams in the GPUs increases with each generation, along with a high available memory bandwidth, memory efficiency is also crucial in order to achieve desired performance. In presence of multiple concurrent data streams, the inherent locality in a single data stream is often lost as these streams are interleaved while moving through multiple levels of memory system. In DRAM based main memory, the poor request locality reduces row-buffer reuse resulting in underutilized and inefficient memory bandwidth. \\\
In this paper we propose Memory-Aware Reordered Source (\textit{MARS}) architecture to address memory inefficiency arising from highly interleaved data streams. The key idea of \textit{MARS} is that with a sufficiently large lookahead before the main memory, data streams can be reordered based on their row-buffer address to regain the lost locality and improve memory efficiency. We show that \textit{MARS} improves achieved memory bandwidth by 11\% for a set of synthetic microbenchmarks. Moreover, MARS does so without any specific knowledge of the memory configuration. 
\end{abstract}

\section{Introduction}

\label{intro}

Last decade has seen an unprecedented growth in data-intensive applications. All computing domains, from mobiles to servers, are increasingly demanding faster, high bandwidth memory, making it critical for overall system performance and computing experience. The demand is particularly high in graphics processors (GPUs) where alongwith traditional 3D games emerging workloads such as vision processing, perceptual computing and machine learning are pushing the limits on memory performance.

GPUs are massively data parallel computing engines with several identical processing elements capable of running multiple threads in a SIMD fashion; this makes GPUs ideal for throughput-oriented tasks where along with exploiting inherent data-level parallelism, high memory latency is hidden by virtue of their parallel architecture. As the number of threads and data streams they operate on increases with each GPU generation, along with the available memory bandwidth, memory efficiency is also critical in order to deliver high performance. While GPU tasks exhibit locality in memory accesses at the software level, the presence of multiple concurrent (and independent) sources of memory requests in the hardware often leads to loss in locality as they reach the memory controller, resulting in poor memory efficiency.

Existing solutions to tackle this include:
 
\begin{itemize}
\item adding more DDR channels or larger caches to provide higher bandwidth---this increases the available bandwidth but at the cost of more power and die area, and the loss in locality is not recovered, making the memory still operate inefficiently
\item increasing memory controller pending queues---this allows a larger lookahead at the memory channel interface, however it also requires increasing the size of all the buffers from GPU all the way to the memory, which increases power and area prohibitively. Secondly, as memory controller is shared typically with multiple CPU cores and GPU, increasing queues will adversely affect CPU performance because of increased memory latency. Hence this option is unpractical. 

\item stream-specific or locality-aware arbitration within GPU, as suggested in ~\cite{ref:alternate_kim} ~\cite{ref:noc_coalsce}, ---this provides marginal benefit since there are multiple arbitration points for different streams and processing elements in the internal interconnection network. Maintaining locality when requests get merged at various locations before they reach the memory is challenging with internal-to-GPU arbitration mechanisms. 

\end{itemize}

In this work, we first show how bad is the locality problem. Then building on the intution of re-ordering memory requests to undo the loss in locality, we propose Memory Aware Reordered Source or MARS, where we show how marginal investment in microarchitecture can significantly boost memory efficiency, thereby improving overall application performance. MARS buffers a large number of memory requests at the GPU boundary and reorders them online to extract locality. This allows the memory controller to operate at close to maximum efficiency for the kinds of streaming applications GPUs promise high acceleration potential. Our simulation results on five memory intensive microbenchmarks show 11\% improvement in memory efficiency by regaining locality using MARS. 

Finally, though we discuss our proposal in context of GPUs in the paper but the architecture and principles used in MARS can be more generally applied in the context of any IP that: a) operates concurrently on multiple streams in heavy throughput mode b) suffers from memory inefficiency due to the IP's inability/insufficiency to control locality while requests exit the IP. MARS would then be implemented at the IP boundary for maximum efficiency boost at the low SoC-level complexity/area/power. 
\section{Motivation}
\label{sec:mot} 

\begin{figure}[h!] 
  \includegraphics[scale=0.5, clip=true, trim=2cm 5cm 2cm 5cm, width=.5\textwidth]{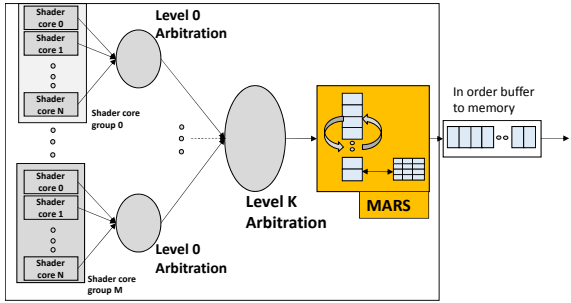}
  \vspace{-15pt}
 \caption{\small An illustrative diagram of the modeled system. MARS is described in Section~\ref{sec:architecture}.}
\label{fig:system}
\end{figure}

We model a GPU with a single fixed-function pipeline and multiple shader cores, clustered into shader core groups, an architecture similar to ~\cite{ref:gpu_jayesh, ref:ivy}. Each shader core has 7 thread contexts, 2 ALU pipelines, 1 load-store pipeline and 1 conditional pipeline. Every cycle one SIMD instruction each from two selected threads are issued within each core. Each shader core can issue up to 1 memory request from the loaded thread contexts per cycle. Our simulator models multiple levels of stream-specific~\footnote{Here streams refer to graphics data typically used in GPUs such as texture, depth, etc.} L1/L2 caches per shader core group and an L3 cache shared by all shader core groups. Memory requests from the shader cores are arbitrated at multiple levels inside shader core groups and finally merged before the L3 cache. All L3 cache misses go off-chip to a buffer from where the requests are forwarded to the memory controller. This is shown in Figure~\ref{fig:system}. 

\begin{figure}[h] 
  \includegraphics[scale=0.5, clip=true, trim=0.2cm 3cm 0.2cm 3cm, width=.5\textwidth]{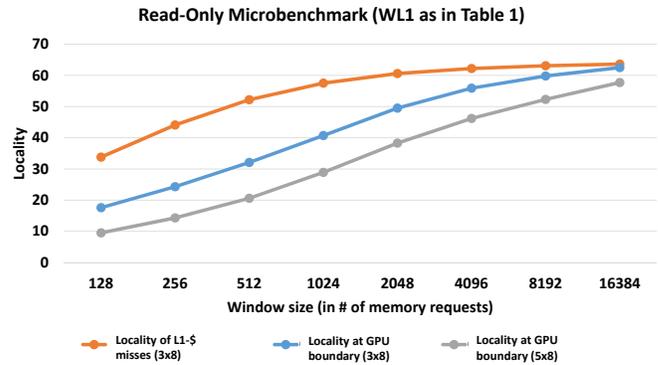}
  \vspace{-15pt}
 \caption{\small Impact on locality as the source of requests are increased}
\label{fig:locality}
\end{figure}

We first define \textit{locality} of a graphics data stream as the average number of memory requests to a unique 4KB page in a window of particular number of requests. To understand the loss in locality in such a heavily-arbitrated architecture, we simulated a synthetic workload with total of 24 shader cores having 3 shader core groups of 8 shader cores each. (This workload is designed to only perform texture reads that always miss in L3, thereby stressing the main memory.)  We model a dual channel LPDDR4-3200 memory system. The DRAM part is 8-way banked with a burst length of eight and 15-15-15 (tCAS-tRCD-tRP) latency parameters. In Figure~\ref{fig:locality} we plot the locality of the texture stream at the output of each L1 texture cache (that is, the misses). This is the average across all texture caches. As we can see, there is significant locality at source in small observation windows and the locality increases as the observation window size is varied from 128 to 16384; this shows that there is temporal and spatial locality in accesses within a single 4KB page since this workload and most of the GPU applications are streaming. 

Figure~\ref{fig:locality} also shows locality of memory requests leaving the GPU towards the main memory. For the same window size, as we go from a individual texture cache to L3 cache, the locality is significantly reduced due to the merging of requests from multiple texture caches, which scatters the locality in individual request streams, consequently reducing memory efficiency. As we go to a higher configuration (40 shader cores), we merge requests from more concurrent data streams and caches, which further reduces the locality as shown in the figure. This microbenchmark contains only a single graphics stream; locality takes a further hit when we run realistic workloads that are operating on multiple graphics streams and generating memory requests simultaneously.

Key observations to be made here are as follows:

\begin{itemize}

\item individual data streams exhibit locality at source; greater locality is observed at larger window sizes

\item locality is scattered as the requests are merged at different arbitration points; this is exacerbated as number of concurrent threads is increased

\end{itemize}

Consequently, applications with streaming memory accesses perform sub-optimally due to inefficient memory utilization. The higher GPU configurations are used for higher throughput, but if the memory efficiency is becoming worse, the full potential is not reached. However, with additional lookahead hardware it should be possible to re-order memory requests to regain the lost memory efficiency.

\section{MARS: Memory-Aware Reordered Source}
\label{sec:architecture}

\subsection{Overview}

In a DRAM first a row is activated (ACT command), followed by one or more column accesses (CAS command(s)) on the open row. Such a memory yields highest efficiency when multiple column accesses are performed on an activated row before it is closed; that is, a high CAS/ACT is desirable for high memory efficiency. Given a stream of memory requests, best memory performance can be achieved by reordering them in order to maximize CAS/ACT. A typical memory controller is designed with read/write pending queues which are used to choose which row to open and issue out-of-order CASes to maximize efficiency. However in a system like a GPU, with many concurrent threads issuing memory requests simultaneously, these queues are not helpful because of the limited size. Simply increasing the size of these queues is not trivial as it can have implications for DRAM timings.

 As we observed in Figure~\ref{fig:locality}, due to interleaving of memory requests from multiple, independent sources, locality is lost before they reach the main memory. A small observation window at the memory controller has fewer requests mapping to a unique page/row, while larger window would allows us to capture several requests that map to the same page/row. To this end, MARS buffers a large number of memory requests in a queue at the GPU interface, tracks which memory requests map to a unique page, and then sends all the requests to the same page back-to-back. This allows us to extract the locality in memory accesses lost in the memory hierarchy. Additionally, MARS can do so without any specific memory configuration details exposed inside the GPU.

\subsection{MARS Policies}

In the memory controller, physical address of a request is translated to DRAM device address using memory map. The memory map defines which bits in the physical address will form channel, rank, bank, row, column etc.., according to the configured memory organization. The memory controller stores requests according to rank/bank/row, monitors the timing parameters and schedules the requests to DRAM trying to maximize the memory efficiency ~\cite{ref:dram}. Ideally, if the exact memory organization is known to MARS, it can always choose the order in which the requests are forwarded to the memory controller. However, exposing the memory map to the remaining architecture is not a viable option since it creates dependency on the memory configuration. 

Instead, we use a simple scheme where in a stream of requests we choose all the requests that map to the same physical page~\footnote{Assuming a 4KB page size, all the higher bits of physical address after discarding lower 12 bits forms physical page} and forward them towards the memory controller. Some of the requests of a physical page may get distributed on different channels or ranks based on the memory map. However the requests served on same rank will have exactly same row address utilizing opened row-buffer and therefore increasing CAS/ACT.      

In the next sub-section, we briefly describe a practical hardware implementation for MARS.

\subsection{Architecture}

There are two key operations involved in MARS, (1) inserting memory requests coming out of the GPU into a queue in order to track their locality, and, (2) selecting and forwarding requests from queue to the memory.

\begin{figure}[h] 
  \includegraphics[scale=0.5, clip=true, trim=0cm 0cm 0cm 0cm, width=.5\textwidth]{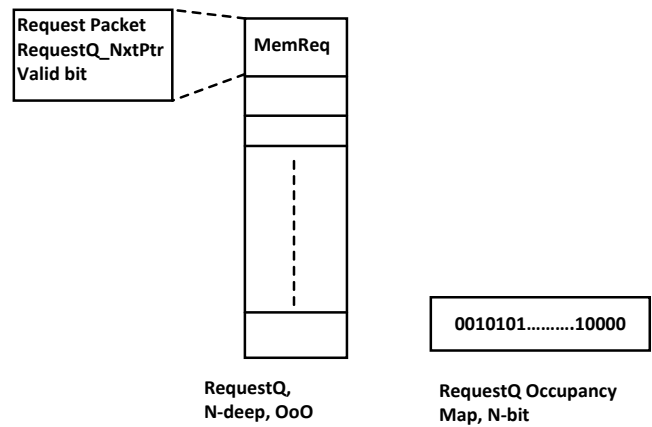}
  \vspace{-15pt}
 \caption{\small RequestQ structure}
\label{fig:requestq}
\end{figure}

\begin{figure}[h] 
  \includegraphics[scale=0.5, clip=true, trim=0cm 0cm 0cm 0cm, width=.5\textwidth]{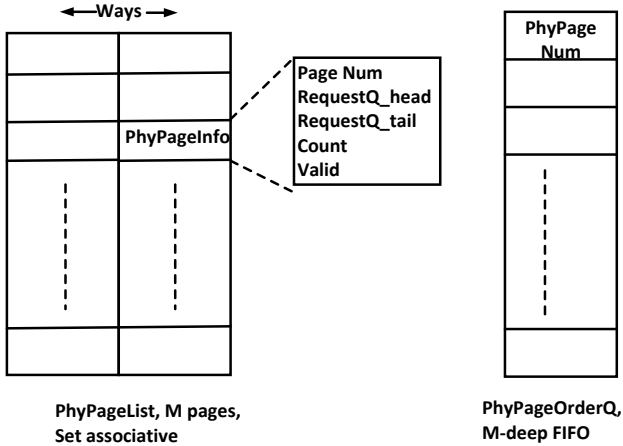}
  \vspace{-15pt}
 \caption{\small PhyPageList and PhyPageOrderQ structures}
\label{fig:pagelist}
\end{figure}

\begin{figure}[h] 
  \includegraphics[scale=0.5, clip=true, trim=0cm 0cm 0cm 0cm, width=.5\textwidth]{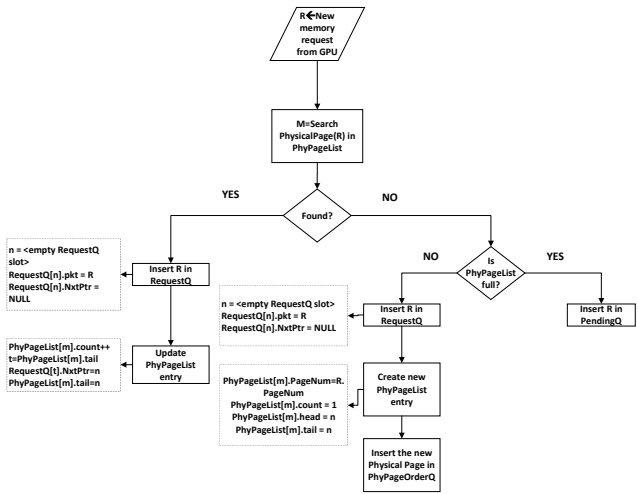}
  \vspace{-15pt}
 \caption{\small Flowchart of insertion algorithm}
\label{fig:insertion}
\end{figure}

\textbf{Insertion}: In MARS, we add a hardware storage structure, called the Request Queue (RequestQ) (shown in Figure~\ref{fig:requestq}), between the GPU ports and the memory controller. Key purpose of RequestQ is to buffer requests in order to provide a large lookahead to capture the locality in the request stream. To implement that, we need to track requests mapping to same 4KB physical page. For this we add another storage structure, Physical Page List (PhyPageList as shown in Figure~\ref{fig:pagelist})), where we create an entry for each unique physical page requests in the RequestQ map to. Each entry in the PhyPageList also stores a list of all the requests in the RequestQ that map to that physical page. To efficiently store this information, we use a linked list organization where each entry in the RequestQ is linked to the chronologically next entry in the RequestQ on the same physical page; this way we need only store the head and tail indices in the PhyPageList to access all the requests belonging to a physical page without needing an associative lookup. In our current implementation, requests can be inserted and extracted from any RequestQ slot; for this we also maintain an occupancy bit-vector map for the RequestQ which is used to find an empty RequestQ slot while inserting a new request. PhyPageList is organized as a set-associative structure, indexed by the physical page number of the memory requests. Figure~\ref{fig:insertion} shows the details of insertion algorithm.

 \begin{figure}[h] 
  \includegraphics[scale=0.5, clip=true, trim=0cm 0cm 0cm 0cm, width=.5\textwidth]{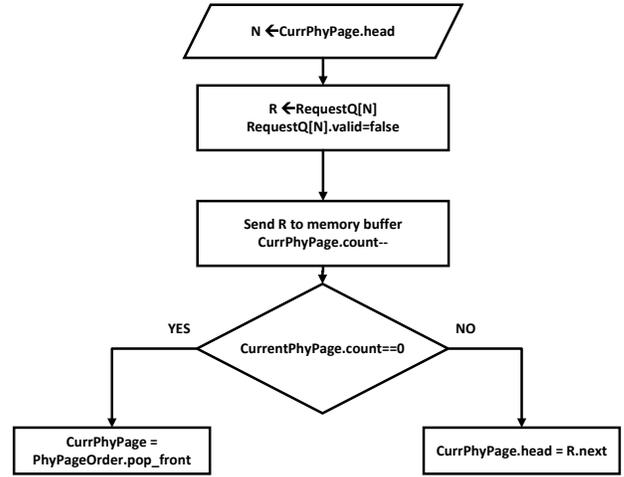}
  \vspace{-15pt}
 \caption{\small Flowchart of forwarding algorithm}
\label{fig:forwarding}
\end{figure}

\textbf{Forwarding a request from RequestQ to MC}: Since MARS sends requests from a single physical page together, we need to first select the physical page we want to forward. For this we use a simple scheme where we always choose a physical page which has the oldest available request in the RequestQ. We add another hardware structure, called Physical Page Order Queue, or PhyPageOrderQ, where we simply store the address of the unique physical page entries in the PhyPageList. This avoids searching for a new page when we have exhausted the current page since PhyPageOrderQ is always chronologically ordered. Figure~\ref{fig:forwarding} shows the details of insertion algorithm.

\textbf{Area Overhead}: In terms of area, we mainly need to account for the hardware storage structures (RequestQ, PhyPageList, PhyPageOrderQ). PhyPageOrderQ is modeled as a simple FIFO, while RequestQ and PhyPageList can be built with SRAMs. Even for a lookahead size of 512 requests over 128 unique physical pages, the area overhead is negligible compared to the GPU area.

\section{Simulation Results}
\label{sec:results}

In this section we provide preliminary results of MARS showing memory bandwidth improvements over a baseline without MARS.

As described in Section~\ref{sec:mot}, we model GPU architecture similar to ~\cite{ref:gpu_jayesh, ref:ivy}. We configure 64 shader cores running at 1GHz. For the main memory, we model LPDDR4-3200 with two single ranked channels. MARS is configured with 512 entry RequestQ and 128 entry 2-way set-associative PhyPageList structure. We simulated 5 synthetic memory intensive benchmarks described in Table~\ref{tab:workload}.

\begin{table}
\begin{center}
\small
\begin{tabular}{|c|l|} \hline
\textbf{Workload} & \textbf{Description} \\ \hline \hline
WL1 & Read only, single texture stream \\ \hline
WL2 & Read + Write, stencil and color streams \\ \hline
WL3 & Write Only, single stream \\ \hline
WL4 & Read Only, HiZ and depth stream \\ \hline
WL5 & Read + Write, single HiZ streams \\ \hline
\end{tabular}
\caption{Simulated workloads}
\label{tab:workload}
\end{center}
\end{table}

\begin{figure}[h] 
  \includegraphics[scale=0.5, clip=true, trim=1cm 5cm 1cm 5cm, width=.5\textwidth]{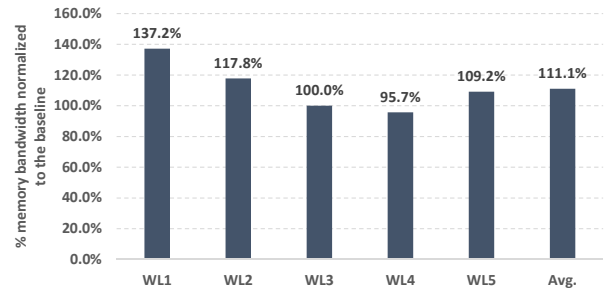}
  \vspace{-15pt}
 \caption{\small Improvement in achieved memory bandwidth}
\label{fig:band}
\end{figure}

\begin{figure}[h] 
  \includegraphics[scale=0.5, clip=true, trim=1cm 5cm 1cm 5cm, width=.5\textwidth]{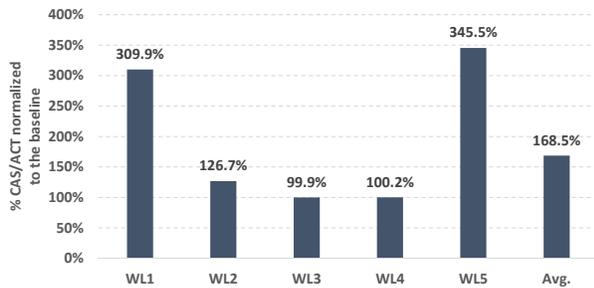}
  \vspace{-15pt}
 \caption{\small Improvement in CAS/ACT}
\label{fig:casperact}
\end{figure}

MARS reorders requests so that the memory controller can efficiently serve more requests (CAS commands) on an open row (ACT). In Figure~\ref{fig:band}, we show percentage improvement in achieved memory bandwidth normalized to the baseline without MARS. Across all the 5 microbenchmarks, MARS improves memory efficiency by 11\% over the baseline. This is primarily due to increase in effective CAS/ACT as shown in Figure~\ref{fig:casperact}. In \textit{WL1} and \textit{WL5}, MARS improves CAS/ACT by more than 2x. Overall, CAS/ACT increases by 69\% over the baseline.

\vspace{-.1in}
\section{Related Work}
\label{sec:related}

Multiple recent efforts ~\cite{ref:moguls,ref:lamar,ref:apres,ref:ccws,ref:dublish,ref:mascar,ref:mrpb,ref:oaws,ref:owl, ref:app_jog, ref:complex_yuan, ref:potential} focus on improving GPU cache and memory efficiency. MARS can be integrated with these solutions and help in improving the overall achieved performance.

Staged Memory Scheduling~\cite{ref:sms} (SMS) targets memory controller design in a multi-core CPU and integrated GPU system. The memory request patterns, bandwidth and latency from GPU and CPU workloads vary drastically and traditional memory controller is not ideal for both memory efficiency and fairness. Therefore, SMS breaks the memory controller design functionally at three higher levels and resolves the issues. However, SMS significantly changes the memory controller design, while MARS is transparent to the memory controller. 

The purpose of Superpackets~\cite{ref:alternate_kim} and Coalescing~\cite{ref:noc_coalsce} concepts is similar to MARS, which is to regain the lost locality in GPU streams. However they give higher priority to locality in arbitration or NOC selecton algorithm. In a GPU system with multiple arbitration levels, reordering requests locally does not yield much benefit as the locality can degrade at the next level, while MARS performs a global reordering.

\vspace{-.1in}
\section{Conclusions}
\label{summary}

Memory bandwidth is crucial in throughput-oriented workloads. However achieving higher memory efficiency in presence of multiple concurrent data streams can be challenging. MARS provides a mechanism to buffer a large number of requests at the boundary of throughput oriented IPs and reorders them to achieve locality thus improving memory efficiency. On 5 memory intensive GPU microbenchmarks, we show that MARS improves achieved memory efficiency by 11\%.

\bibliographystyle{IEEEtranS}
\def\REF{\bibitem}

\end{document}